\documentstyle[seceq,epsf]{ptptex}

\title{
Universal behaviors in granular flows  and traffic flows
}

\author{
Hisao {\sc Hayakawa}\footnote{email: hisao@phys.h.kyoto-u.ac.jp
}
and Ken {\sc Nakanishi}$^{*,}$\footnote{email: tmknaka@eng.shizuoka.ac.jp
}
}

\inst{
Graduate School of Human and Environmental Studies,
Kyoto University, Kyoto 606-01, Japan\\
$^*$ Department of Mechanical Engineering, Shizuoka University,
Hamamatsu 432, Japan
}

\recdate{ November 8, 1997
}

\abst{
We review the current understanding on
universal behaviors in granular flows through a vertical 
pipe and traffic flows.
We carry out weakly nonlinear analysis 
of a  model for traffic flows based on the technique of
soliton perturbations, and determine the selected propagating
velocity, the amplitude, the width of interfaces connecting between
jam phase and non-jam phase.
From the direct simulation of the model,
we have confirmed the validity of our theoretical analysis.
We also introduce a  model for granular pipe flow
supplemented by the white noise, which reproduces
$P(f)\sim f^{-4/3}$, where
$P(f)$ is the  power spectrum in the frequency $f$. 
(This paper will be published in Progress of  Thoeretical Physics Supplment).
}

\begin{document}

\maketitle

\section{Introduction}

Recently,  cooperative dynamics in
 dissipative systems 
consisting of discrete elements
have attracted much attention.
Researches on granular materials are
 efforts to understand unusual behaviors of
discrete element systems\cite{granules}
such as convection\cite{taguchi}, size
segregation\cite{rosato87}, bubbling\cite{pak94}, standing waves
\cite{swinney95} and localized excitations\cite{swinney96}
as well as thermodynamic descriptions of granular particles
under the vertical vibrations\cite{thermo}.
In particular, it is interesting that 
jam formation of particles  and a power law in the power spectra
in flows of granular particles 
through a narrow  vertical pipe.
Similarly, 
 traffic jams in a highway is also  an attractive subject 
 not only for engineers but for physicists\cite{traffic}.
Similarities between two phenomena are obvious.
Both consists of discrete dissipative elements, vehicles and particles
which are confined in a quasi one-dimensional systems
such as a highway and a pipe.
There is an optimal velocity in each system,
the competition between  the relaxation to the optimal velocity 
and acceleration of particles produces jam formation.
We, thus,  expect that there  exists  common and universal 
mathematical structure behind these phenomena.

In the next section, we will introduce 
typical models for granular flows  and 
traffic flows, and summarize 
the current status of our understanding on
universal properties in one-dimensional flows.
Among these studies we will focus on
 recent two main streams  
in studies of granular flows  and traffic flows.
First, we will review  the recent 
progress in theoretical analysis of pure one-dimensional models.
The second, we will introduce the progress  in analysis
for power spectra in density auto-correlation function
in quasi one-dimensional systems.

From the theoretical analysis of  pure one-dimensional models, at least,
 it has been confirmed that 
 perturbed  solitons by dissipative corrections play important roles
in traffic flows and granular flows.
In particular, Komatsu and Sasa\cite{komatsu} have revealed 
the mechanism of jam formation in a traffic flow by 
the perturbative treatment of solitons.
Hayakawa and Nakanishi\cite{ken} have generalized the analysis of
Komatsu and Sasa\cite{komatsu}, and demonstrate that universal mathematical
structure exists in pure one-dimensional models for granular flows 
and traffic flows. In section 3 consisting of three subsections, we
will review the details of theoretical argument by Hayakawa and
Nakanishi\cite{ken} which discussed a model of traffic flow.
We will stress that the framework of our analysis can be used in any
one-dimensional models for granular flows and traffic flows.

  In realistic situations, however, e.g. highways
 have several lanes, and vehicles (particles) can pass  
slow vehicles (particles). 
When we include multi-lane effects in one-dimensional models,
separations between jam and non-jam phases become obscured.
 However, there is another universal law
in quasi one-dimensional systems for flows of dissipative discrete elements,
i.e. a power law in power spectrum of density auto-correlation
function of vehicles or 
particles. Recently, Moriyama et al.\cite{chuo}
have presented  that the power spectrum is given by
 $P(f) \sim f^{-\alpha}$ with 
 $\alpha \cong 1.33$ from their experiment on 
granular flow through a vertical pipe.  This result is 
expected to be
a universal  in quasi one-dimensional dissipative flows such as
traffic flows in a highway.
Thus, we will clarify the mechanism to appear $f^{-4/3}$ law in section 4.
For this purpose, we will introduce a simple model 
 supplemented by the
white noise. From the simulation of our model we will confirm 
that our model can reproduce $f^{-4/3}$. We will explain the mechanism 
how to obtain $f^{-4/3}$-law from the simple analytic calculation.

In section 5, we will give concluding remarks, especially on
 universality 
in granular flows through a pipe
and traffic flows. We also summarize our results.

\section{Models}

In this section let us summarize what models exist
and what consensus in  studies of granular flows  and traffic flows
is obtained.

There are many models to describe traffic flows and granular flows
through a pipe.
We believe that universal behaviors do not depend on the choice of
the model. 
Recently, Hayakawa and Nakanishi\cite{ken}
have proposed  a generalized optimal velocity model for
traffic flows
\begin{equation}\label{rOV}
 \ddot x_n=a[U(x_{n+1}-x_n)V(x_{n}-x_{n-1})-\dot x_n],
\end{equation} 
where $x_n$ and $a$ are the positions of $n$ th car, and 
the drag coefficient, respectively. 
This model contains  the psychological effect of drivers.
Namely,  the driver of $x_n$ takes care of not only the distance
ahead $x_{n+1}-x_n$ but also the  distance behind 
 $x_n-x_{n-1}$. 
The optimal velocity function $U$ should be a monotonic increasing 
function of the distance of $x_{n+1}-x_n$ and $V$
 should be a monotonic decreasing function
of $x_n-x_{n-1}$. Thus,  we adopt
\begin{equation}\label{back}
U(h)=\tanh(h-2)+\tanh(2); \quad  V(h)=1+f_0(1-\tanh(h-2))
\end{equation}
for the explicit calculation in section 3, where $f_0$ is a constant.
We put these optimal velocity 
functions as the product form $UV$ in (\ref{rOV}),
 because the driver of $x_n$ cannot accelerate the car without enough
space ahead  even when the distance $x_n-x_{n-1}$
becomes short.
In other words, the model including $U(x_{n+1}-x_n)+V(x_n-x_{n-1})$ 
is mathematically unstable and unphysical, because the acceleration
by $V$ causes crash of vehicles. 
This model (\ref{rOV}) with (\ref{back}) is a generalization of
the optimal velocity (OV) model proposed by Bando et al. \cite{bando}
\begin{equation}\label{OV}
\ddot x_n=a[U(x_{n+1}-x_n)-\dot x_n].
\end{equation} 

The generalized OV model is similar to the model
 of granular flow in a one-dimensional tube
\begin{equation}\label{powder}
\ddot x_n=\zeta[\tilde U(x_{n+1}-x_{n-1})-\dot x_n]+T[\varphi'(x_{n+1}-x_n)
-\varphi'(x_n-x_{n-1})],
\end{equation}
where $\zeta$ and $T$ are respectively the drag coefficient
and the strength of collision among particles.
$\tilde U$ and $\varphi$ are the optimal velocity 
and soft core repulsion potential, respectively\cite{chuo}.

There is a fluid field model to describe traffic flows\cite{KK} which 
consists of mass conservation
for density field $\rho$ and momentum conservation  for 
velocity field $v$ as
\begin{eqnarray}
\label{KK}
\partial_t \rho &=& -\partial_x (\rho v), \nonumber \\
\partial_t v &=& -v \partial_x v -\frac{T_e}{\rho} \partial_x \rho
+\frac{U_{\rho}(\rho)-v}{\tau_{\rho}}+\frac{\eta}{\rho}\partial_x^2 v,
\end{eqnarray}
where $\tau_{\rho}$ and $T_{e}$ are 
a characteristic time for the relaxation and the effective temperature,
respectively. This model also contains the relaxation 
mechanism to an optimal
velocity $U_{\rho}$, while the pressure term and the viscous term
are phenomenologically introduced to stabilize the solution of a set 
of equations (\ref{KK}).

This fluid model (\ref{KK}) is similar to fluid models to describe granular
 flows through a pipe
and fluidized beds\cite{batchelor,sasa,komatsu93,goez} and 
mixture of polymers\cite{doi}. As a typical example, we write an
explicit form of a fluid model for granular flow by Sasa and 
Hayakawa\cite{sasa} at the Froude number $Fr$:
\begin{eqnarray}\label{sasa}
\partial_t \rho&=-&\partial_x (\rho v), \nonumber \\
\partial_t v &=& -v\partial_x v -\tilde\zeta(\rho)(V_0-v)-\frac{1}{Fr}
               -f''(\rho)\partial_x \rho +\kappa \partial_x^3 \rho 
\nonumber \\
   & &           +\frac{1}{\rho}\partial_x(\rho \nu(\rho)\partial_x v).
\end{eqnarray}
Although this model seems to be complicated, 
 each term corresponds to that in  eq.(\ref{KK}).
Here the term in proportion to $\kappa$ is introduced to stabilize
the solution furthermore, which can be regarded as a coupling term in
mixing free energy. The pressure $f'(\rho)$ arises from the collisions
among particles, which is a recovery force to the mean density $\rho_0$
and diverges at the closest packing states at $\rho_{cp}$.
The kinetic viscosity $\nu(\rho)$  also diverges at $\rho_{cp}$.
Komatsu and Hayakawa\cite{komatsu93}, thus,  adopted
$f(\rho)=\beta_1 (\rho-\rho_0)^2/(\rho_{cp}-\rho)$
and $\nu(\rho)=\beta_2/(\rho_{cp}-\rho)$ for the simulation of
(\ref{sasa}), where $\beta_1$ and $\beta_2$ are constants.
The optimal velocity $V_0$  is assumed to be
 a constant because of the incompressibility 
of the fluid, and $\tilde\zeta(\rho)$ is a phenomenological function
to be determined by the sedimentation rate.

At first sight fluid models (\ref{KK}) and (\ref{sasa}) 
are very different from 
discrete models such as (\ref{rOV}), (\ref{OV}) and (\ref{powder}).
However, there is  common mathematical structure.
The fluid models of granular flows such as eq.(\ref{sasa})
 are   reduced to the Korteweg-de Vries
(KdV) equation 
near the neutral curve 
of the linear stability\cite{sasa,komatsu93,goez}.   
Kurtze and Hong\cite{kurtze} also derived KdV equation from the fluid model
of the traffic flow (\ref{KK})\cite{KK}.  
Of course, it is easy to derive KdV equation
from the discrete models (\ref{rOV}),
(\ref{OV}) and (\ref{powder})  near the neutral
curve. Thus, at least, we get a consensus that dissipative solitons
play important roles  for granular flows  and
traffic flows.

Unfortunately KdV equation is not adequate to
describe the traffic jams, because  solutions of these equations 
are essentially pulses and no interface solutions  connecting jam phase with
non-jam phase are included.
Komatsu and Sasa\cite{komatsu} 
solved such a puzzle from the analysis of  the original
 OV model (\ref{OV}).
They have showed that 
(\ref{OV}) can be 
reduced to the modified KdV (MKdV) equation 
at the critical point (the averaged car distance $h=2$)
or the most unstable point on the neutral curve.
They also show that   
symmetric kink solitons deformed by
dissipative corrections describe a phase separation between 
bistable phases. 
This analysis is also consistent with recent 
analysis for
the exactly solvable models  which may be regarded as 
simplified  optimal velocity models\cite{exact}. 
 However, as will be shown, 
the generalized optimal  velocity model (\ref{rOV}) and 
granular model (\ref{powder}) as well as
 the fluid model (\ref{KK}) of traffic flows\cite{KK} 
and  fluid models, e.g. (\ref{sasa}) 
for granular flows\cite{sasa,komatsu93,goez}  
are not reduced to MKdV equation  at 
the critical point or the most unstable point on the neutral curve.
In fact, Komatsu\cite{komatsu2} has shown that
the fluid model (\ref{KK}) has the following properties:
(i) Interfaces (kinks) between jam and non-jam phases are asymmetric.
(ii) The critical point to appear kinks is, in general, 
different from the most unstable point on the neutral curve.
 (iii) Eventually one branch of the coexistence curve
 exists in the linearly unstable region.
He also demonstrates that 
MKdV equation is recovered in 
a special choice of parameters of the fluid model,
while fluid models cannot be reduced to MKdV equation in general cases.
Thus, we need to clarify universal 
characteristics of dissipative particle dynamics
in general cases which contains (\ref{rOV}), (\ref{powder}) and 
fluid models\cite{KK,batchelor,sasa,komatsu93,goez}.  
For this purpose, we will focus on analysis the simplest model (\ref{rOV}) 
among them to characterize the phase separation between jam and
non-jam phases.

We may  be suspicious of reality and relevancy 
of pure one-dimensional models.
As indicated in Introduction,
when we include multi-lane effects in one-dimensional models,
separations between jam and non-jam phases become obscured.
However, the fundamental characteristics of pure one-dimensional models
should be important for quasi one-dimensional cases.
To demonstrate the relevancy of pure one-dimensional models, 
we will focus on another universal law
in quasi one-dimensional systems,
i.e. a power law in power spectrum of density auto-correlation
function of vehicles or 
particles.

Power-law form of the power 
spectrum $P(f) \sim f^{-\alpha}$, where $f$ is frequency, of density 
fluctuations was  found in both numerical simulations\cite{peng94} and 
experiments\cite{horikawa95,horikawa96}.  Although their interpretations 
on the origin of the emergence of density waves are different, estimated 
values of the exponent $\alpha$ is close to each other 
($1.3 < \alpha < 1.5$).
  The previous reports\cite{horikawa95,horikawa96} on the estimation 
$\alpha \cong 1.5$ seem a little 
ambiguous: 
The volume of air flow out of the 
bottom end of the pipe was not well controlled.  Besides, the power spectra 
they obtained were still noisy.  
Recently, Moriyama et al.\cite{chuo} 
have  presented 
better-controlled 
air flow out of the pipe and more accurate experimental results than the 
previous ones by increasing the number of trials.  
One of their results is the precise estimation of 
the scaling exponent of the power spectrum $P(f) \sim f^{-\alpha}$
with  $\alpha \cong 1.33$.  This result is identical to
that by LGA\cite{peng94}, and is expected to be
a universal law in quasi one-dimensional dissipative flows such as
traffic flows in a highway.

The second  purpose of this paper
 is to clarify the mechanism to appear 
$f^{-4/3}$ law in power spectra.
We, thus,  extend the one-dimensional 
model  (\ref{powder}) 
to a stochastic model supplemented by 
the white noise  as
\begin{eqnarray} \label{1d_powder}
\ddot h_{n} &=&  \zeta [
\tilde U(\frac{h_{n+1}+h_n}{2})-
\tilde U(\frac{h_n+h_{n-1}}{2})] -\dot h_n] \nonumber \\ 
& & +T[\varphi'(h_{n+1})+\varphi'(h_{n-1})-2\varphi'(h_{n})]
 + f_{n}(t),
\end{eqnarray}
where $h_{n}=x_{n+1}-x_n$.
The most crucial simplification of the model
is that  $f_n$ is assumed to be the Gaussian white noise with zero mean. 
  The optimal velocity 
$\tilde U(h)$   may be the sedimentation rate which is a function of the local 
volume fraction\cite{sedi} in general. 
 It should be noticed that 
the  drag $\zeta$ is irrelevant in  systems
where the bottom end of the pipe is fully open,
because air in the pipe flows away together with particles.
Thus, to observe density waves it is important 
to close the cock of the pipe. 

One of the important points is that the white noise is introduced 
not to an equation for $x_n$ but
 to an equation for the relative motion of particles $h_n$, because
the ordering of particles along a pipe or a  highway  is
not conserved due to passing of particles in multi-lane or
multi-dimensional systems.
From both simulation and analytic calculation of (\ref{1d_powder}),
we will demonstrate that this simple model reproduces $\alpha = 4/3$  near the 
neutral curve of the linear stability analysis of uniform states
in section 3.
 We can expect the existence of a universal law regardless to the choice
of a specific model.
Therefore, the model (\ref{rOV}) for $h_n$ supplemented by the white noise
is expected to 
belong to the same universality class that by (\ref{1d_powder}).
The fluid models (\ref{KK}) and (\ref{sasa}) 
with non-conserved white noise  also should
behave  similarly as those for discrete models. 

Now, we have come back to an interesting question: What are good models
among many models? The answer is simple.
The simple models are good ones.
For example, when we compare discrete models (\ref{rOV}) and (\ref{powder}) 
 with fluid models (\ref{KK}) and (\ref{sasa}), the representation of
discrete models are  shorter than fluid models.
Since fluid models are partial differential equations,
we may need careful check of the validity of the simulation scheme
and long CPU time to simulate them.
On the other hand, discrete models which are coupled ordinary 
differential equations
are free from  such problems.
The theoretical analysis for discrete models is also simpler than
that for fluid models, because each discrete model contains only one 
set of equations.
On the other hand, eq.(\ref{OV}) is a oversimplified model which loses
some of universal properties.
Therefore, we believe that models (\ref{rOV}) and (\ref{powder}) are
the most fundamental ones. 
In the following sections, thus,
we will focus on the analysis of discrete
models (\ref{rOV}) and (\ref{1d_powder}).

\section{Theory of a pure one dimensional model for traffic flow}

In this section, we concentrate on the analysis of (\ref{rOV}).
The result explained in
this section is universal for all of models introduced in the previous
section.
This section consists of three subsections.
In the first subsection, we will summarize the result of
linear stability analysis for uniform flows. In the next subsection,
the main part of this section, the details of weakly nonlinear analysis
will be explained, where a steady propagating solution is presented
and the selection of its propagating velocity, the width of interfaces
and the amplitude will be discussed.
In the last subsection, we will confirm the validity of our theoretical
analysis from the comparison between
 direct simulation and theoretical results.  

\subsection{ Linear stability of uniform flow}

In this subsection we summarize the linear stability analysis
of the uniform propagating flow. 

It is obvious that
there is a constant propagating solution with $x_{n+1}-x_n=constant$. 
Let us rewrite (\ref{rOV})  as
\begin{equation}\label{rOV2}
\ddot r_n=a[U(h+r_{n+1})V(h+r_n)-U(h+r_n)V(h+r_{n-1})-\dot r_n],
\end{equation}
where   $h$ is the averaged distance of successive cars
 and $r_n$ is $x_{n+1}-x_n-h$ . 
The linearized equation of (\ref{rOV2}) around $r_n(t)=0$
 is given by
\begin{equation}\label{linear} 
\ddot  r_n=a[U'(h)V(h)( r_{n+1}- r_n)
+U(h)V'(h)( r_n- r_{n-1})-\dot  r_n],
\end{equation}
where  the prime refers to the differentiation 
with respect to the argument. 
 With the aid of the Fourier transform
\begin{equation}
r_q(t)=\frac{1}{N}\sum_{n=1}^N \exp[-i q n h]  r_n(t)
\end{equation}
with $q=2\pi m/N h$ and the total number of cars $N$ 
we can rewrite (\ref{linear}) as
\begin{equation}\label{linearF}
(\partial_t-\sigma_+(q))(\partial_t-\sigma_-(q))r_q(t)=0
\end{equation}
with
\begin{equation}\label{sigmapm}
\sigma_{\pm}(q)=-\frac{a}{2} 
\pm \sqrt{ (a/2)^2 -a D_h[U,V]( 1-\cos(q h) )+
i a (UV)'\sin(q h)} ,
\end{equation} 
where  we drop the argument $h$ in $U$ and $V$.
 $D_h[U,V]\equiv U'(h)V(h)-U(h)V'(h)$ denotes Hirota's derivative.
The solution of the initial value problem in 
(\ref{linearF}) is the linear combination of terms in proportion to 
$\exp[\sigma_+(q) t]$ and $\exp[\sigma_-(q) t]$.
The mode in proportion to  $\exp[\sigma_-(q) t]$
can be interpreted as  the fast decaying mode, while the term in 
proportion to   $\exp[\sigma_+(q) t]$ is the slow and more important  mode.

The violation of the linear
stability of the uniform solution in (\ref{linear}) is equivalent to 
$Re[\sigma_+(q)]\ge 0$. Assuming $qh\ne 0$ ($qh=0$ is the neutral mode),
the instability condition is given by
\begin{equation}\label{neutral-con}
2 (UV)'^2\cos^2(\frac{q h}{2})\ge a D_h[U,V].
\end{equation}
Thus, the most unstable mode exists at $q h\to 0$ and the neutral curve for 
long wave instability is given by
\begin{equation}\label{neutral}
a=a_n(h)\equiv \frac{2(UV)'^2}{D_h[U,V]} .
\end{equation}
The neutral curve 
in the parameter space $(a,h)$ is shown in Fig.1 
for  $f_0=1/(1+\tanh(2))$ in (\ref{back}). 
For later convenience, we write the explicit form of 
the long wave expansion of $\sigma_+$
in the vicinity of the neutral curve 
\begin{equation}
\sigma_+(q)= 
i c_0q h-c_0^2\frac{a-a_n( h)}{a_n( h)^2}(qh)^2
-i\frac{(q h)^3}{6}c_0-
\frac{(qh)^4}{4a_n( h)}c_0^2+O((qh)^5) ,
\label{dispersion}
\end{equation}
where $c_0=(UV)'$.
Thus, the uniform state becomes 
unstable due to the negative diffusion constant  for $a<a_n(h)$.

\subsection{Nonlinear analysis}

The simplest way to describe nonlinear dynamics is the long wave expansion
with the help of a suitable scaling ansatz.
It is easy to derive the KdV equation near the neutral curve 
from (\ref{rOV}) 
as in the case of fluid models\cite{sasa,komatsu93,goez,kurtze}.
 As mentioned in Introduction, 
to describe the phase separations, however,  we should  choose
the critical point $(a,h)=(a_c,h_c)$ at 
$(U(h)V(h))''=0$ where the coefficient of $\partial_x r^2$ becomes 
zero on the neutral curve.
At this point
the cubic nonlinear terms can produce the interface solution
to connect two separated domains.
 The explicit critical point of (\ref{back}) with
$f_0=1/(1+\tanh(2))$  is given by
\begin{equation}\label{cp}
h_c= 2-\tanh^{-1}(1/3)\simeq 1.65343; \quad 
a_c=\frac{512}{81}f_0^2\simeq 1.63866. 
\end{equation}
Unfortunately, the reduced equation based 
on the long wave expansion of 
our model is an ill-posed equation. In fact, 
 the scaling 
of variables as $ r_n(t)=\epsilon
 r(z,\tau)$, $z=\epsilon
(n+c_0t)$ and $\tau=\epsilon^3 t$ with $\epsilon=\sqrt{(a_c-a)/a_c}$
leads to 
\begin{equation}\label{long}
\partial_{\tau}r=a_1\partial_z r^3-a_2\partial_z^3r+a_3
\partial_z^2r^2
\end{equation}
in the lowest order, where
  $a_1$, $a_2$ and $a_3$ are constants.
The solution of (\ref{long}) is blown up within finite time.
The reason is simple. Its linearized equation
around $r=d_0$ is unstable for all scale, because
the solution with $r-d_0\simeq \exp[ikz+\lambda_k \tau]$ 
has the growth rate 
 $Re[\lambda_k]=2  k^2 a_3 d_0$ which is always positive when $a_3 d_0>0$.  
Therefore, the simple long wave expansion
  adopted by Komatsu and Sasa\cite{komatsu} 
for (\ref{OV}) and $a_3=0$  cannot be used in our case.

This short scale instability in (\ref{long}) 
arises from the long wave expansion.
To avoid such the difficulty,
  we only focus on a steady propagating solution
for the theoretical analysis.
Dynamical behavior is followed by the direct simulation of 
discrete model (\ref{rOV}),  where simulation of (\ref{rOV}) is
much  easier than that of 
the reduced partial differential equations such as MKdV.
In addition, our long time simulation suggests that the solution of
(\ref{rOV}) seems to be relaxed to a steady propagating mode.

To obtain the steady 
propagating kink solution, at first
  we  eliminate the fast decaying mode
in (\ref{rOV2}) 
as
\begin{equation}\label{lattice}
(\partial_t -\sigma_+(\partial_x))r(x,t)=(\sigma_+-\sigma_-)^{-1}N[r(x,t)] ,
\end{equation}
where $N[r]$ represents the nonlinear terms 
\begin{eqnarray}\label{def-N}
\frac{N[r]}{a}&=&U(h+e^{h\partial_x}r)V(h+r)
-U(h+r)V(h+e^{-h\partial_x}r) \nonumber \\
& &-U'(h)V(h)(e^{h\partial_x}-1)r(x,t)+U(h)V'(h)(1-e^{-h\partial_x})r(x,t).
\end{eqnarray}
Since $(\sigma_+-\sigma_-)^{-1}$ is the inverse of the polynomial of 
the differential operators, it is convenient to use the 
expansion $(\sigma_+-\sigma_-)^{-1}\simeq 
a^{-1}[1-\displaystyle\frac{2 h}{a}(UV)'\partial_x
+O(h^2)]$.
It should be noticed that  (\ref{lattice}) contains most of
important information and no instability in time evolution.

To obtain the scaled propagating kink solution 
we assume the scaling of the variables by $\epsilon=\sqrt{(a_c-a)/a_c}$ as
\begin{equation}\label{scaling}
r(x,t)=\epsilon \displaystyle\sqrt{\frac{6 \gamma c_0}{|(UV)'''|}}
R(z): \quad z=\epsilon \displaystyle\sqrt{6\gamma}
(n+c_0t-\epsilon^2 \gamma(t) t),
\end{equation}
where the arguments in $U$ and $V$ are fixed at $h=h_c$,
 and $\gamma$ is the positive free parameter which will be determined 
from the perturbation analysis.
The expansion of  $N[r]$ is given by 
\begin{equation}\label{N[r]}
N[r]/a= \sum_{n=1}^{\infty}\sum_{m=2}^{\infty} h^mC_{mn} \partial_x^n r^m
- h^3 U'V'\partial_x r\partial_x^2 r+\cdots ,
\end{equation}
where $C_{21}=\frac{1}{2}(UV)'',  C_{22}=\frac{1}{4}D_h[U,V]', 
C_{23}=\frac{1}{12} (UV)'' , 
C_{31}=\frac{1}{6}(UV)''',  C_{32}=\frac{1}{12}D_h[U,V]'', 
C_{41}=\frac{1}{24}(UV)''''$ with $D_h[U,V]'=
\displaystyle\frac{d}{dh}D_h[U,V]$.
Substituting (\ref{scaling})
into (\ref{lattice}) with the help of (\ref{N[r]})
  we obtain
\begin{equation}\label{steady}
\frac{d}{dz}\left\{\frac{d^2R}{dz^2}- R(R^2-1)+\beta \frac{d}{dz}(R^2)\right\}=
\epsilon\displaystyle\frac{d}{dz}M[R] ,
\end{equation}
where $\beta=3D_h[U,V]'/(2\sqrt{c_0|(UV)'''|})$
 and   
\begin{equation}\label{M}
M[R]=\sqrt{\gamma}[\rho_{23}\left(\frac{d R}{dz}\right)^2-
\rho_{32}\frac{dR^3}{dz}-\rho_{41}R^4
-\frac{1}{4\eta}(4 \frac{dR}{dz}+\frac{d^3R}{dz^3}-
\frac{2}{\gamma}\frac{dR}{dz})]+
\dot \gamma[ \frac{z R}{2\gamma^{5/2}}-
\gamma t R] .
\end{equation}
Here $\dot\gamma=d\gamma/d\tau$ with $\tau=\epsilon t$,
and  $1/\eta=\sqrt{6}D_h[U,V]/c_0$,
$\rho_{23}=3 \sqrt{6}U'V'/\sqrt{c_0|(UV)'''|}$,
$\rho_{32}=\sqrt{3/2}D_h[U,V]''/|(UV)'''|$ and 
$\rho_{41}=\sqrt{3c_0}(UV)''''/(2\sqrt{2|(UV)'''|^3})$.

Assuming $R(z)=R_0(z)+\epsilon R_1(z)+\cdots$,
 we obtain a
solution
\begin{equation}\label{asymmetry}
R_0^{(\pm)}(z)=\tanh(\theta_{\pm}z); \quad \theta_{\pm}=
\frac{\beta\pm \sqrt{\beta^2+2}}{2}
\end{equation}
in the lowest order.
This solution represents a kink or an anti-kink
connecting between jam and non-jam phases because
of $\theta_+>0$ and $\theta_-<0$.
Notice that the solution is not localized one and does not
satisfy the periodic boundary condition.
Therefore, we need a careful  treatment of the boundary condition. 
In fact, our preliminary result\cite{ken2} suggests that
the selected value of $\gamma$ under the open boundary condition 
is different from that under the periodic boundary condition.  

In this paper we restrict ourselves to the case under the periodic boundary.
To satisfy the periodic boundary condition, we use
\begin{equation}\label{zero-pbc}
R_0(z)\simeq {R_0}^{(+)}(z-z_+)-1+{R_0}^{(-)}(z-z_-)
\end{equation}
as an approximate solution of the lowest order equation (\ref{steady}),
where a pair of kink and anti-kink exists at $z=z_+$ and $z=z_-$. 
Since (\ref{zero-pbc}) is not an exact solution of (\ref{steady}),
there should be an interaction between the kink and the anti-kink
which is an exponential function of the  distance between 
them\cite{ken2}.

Now, let us discuss the effect of perturbative terms in (\ref{steady}).
It is known that perturbation of solitons or solution including
a free parameter  becomes unstable except for
the solution where the parameter has
a  special value.\cite{komatsu,hinch,ott,kivshar,needs,ei,chen}.
The linearized equation of (\ref{steady}) can be reduced to
\begin{equation}\label{linearized}
{\cal L} R_1=\frac{d}{dz}
 M[R_0] ,
\end{equation}
where
\begin{equation}\label{L-op}
{\cal L}=\partial_z^3+\partial_z-6R_0\partial_z-3R_0^2\partial_z
+2\beta \partial_z^2R_0+4\beta \partial_zR_0\partial_z+2\beta R_0\partial_z^2 .
\end{equation}
To obtain a regular behavior of perturbation in $O(\epsilon)$ 
the perturbed solution should satisfy the  
solvability condition 
\begin{equation}\label{solva}
(\Psi_0,\frac{d}{dz}M[R_0])\equiv 
\lim_{L\to \infty}\int_{-L}^{L} dz \Psi_0 \frac{d}{dz}M[R_0]=0,
\end{equation}
where $L$ is the system size and  $\Psi_0$ satisfies
\begin{equation}\label{zero-eq}
{\cal L}^{\dagger}\Psi_0=0: \quad 
{\cal L}^{\dagger}=-\partial_z^3-\partial_z+3 R_0^2\partial_z+
2\beta R_0\partial_z^2.
\end{equation}
When we adopt (\ref{zero-pbc}) as $R_0$,
$\Psi_0$ also should satisfies the periodic boundary condition.
Thus, we assume
\begin{equation}\label{psi0}
\Psi_0(z)={\Psi_0}^{(+)}(z-z_+)-1+{\Psi_0}^{(-)}(z-z_-),
\end{equation}
where $\Psi_0^{(\pm)}$ is the solution of (\ref{zero-eq}) 
when we replace $R_0$ by $R_0^{(\pm)}$.
Since $\Phi_0\equiv \partial_z \Psi_0$ satisfies
\begin{equation}\label{phi_0}
\tilde{\cal L}^{\dagger}\Phi_0(z)=0; \quad 
\tilde{\cal L}^{\dagger}=-\partial_z^2-1+3 R_0^2+
2\beta R_0\partial_z,
\end{equation}
the solution of (\ref{zero-eq}) can be expressed by
\begin{equation}\label{psi}
{\Psi_0}^{(\pm)}(z)=\frac{\alpha_{\pm}}{2}\int_{-z}^zdz' 
({\rm sech}[\theta_{\pm} z'])^{1/\theta_{\pm}^2}; \quad
\Psi_0^{(\pm)}(z)
=-\Psi_0^{(\pm)}(-z),
\end{equation}
where we use 
\begin{equation}\label{phi0}
\Phi_0^{(\pm)}(z)=({\rm sech}[\theta_{\pm} z])^{1/\theta_{\pm}^2}.
\end{equation}
The constant $\alpha_{\pm}$ in (\ref{psi}) 
 is determined to satisfy $\Psi_0(\pm \infty)=-1$.
Thus, we obtain
\begin{equation}\label{alpha}
\alpha_{\pm}=\frac{2\theta_{\pm}}{{I_0}^{(\pm)} }
;\quad
 I_n^{(\pm)}=\int_{-\infty}^{\infty} dx ({\rm sech} x)^{1/\theta_{\pm}^2+2n}
=\sqrt{\pi}
\displaystyle\frac
{\Gamma(1/(2\theta_{\pm}^2)+n)}{\Gamma(1/(2\theta_{\pm}^2)+n+1/2)},
\end{equation}
where $\Gamma(x)$ is the gamma function.
It should be noticed that $\Psi_0(z)$ is not a localized function.
Therefore, we cannot neglect the boundary effects in 
the solvability condition (\ref{solva}).  

Let us rewrite (\ref{solva}) as
\begin{equation}\label{solva2}
[\Psi_0 M[R_0]]_{-L}^{L}=(\Phi_0(z), M[R_0]) ,
\end{equation}
where $[f(z)]_{-L}^{L}=f(L)-f(-L)$.
From (\ref{M}) it is obvious that
contribution  from terms except for those in proportion to $\rho_{41}$ and
$ z R$ is zero in the left hand side of (\ref{solva2}) 
under  any boundary conditions. 
Notice that the contribution from 
the term in proportion to $t R$ vanishes because of
its symmetry. If we adopt the periodic boundary condition
and use (\ref{zero-pbc}) and (\ref{psi0}),
the contribution from the term $\rho_{41}$ is canceled\cite{comment}.
Thus, the left hand side of (\ref{solva2}) is reduced to
 \begin{equation}\label{b-eff}
[\Psi_0 M[R_0]]_{-L}^{L}=\frac{\dot \gamma}{\gamma^{5/2}}L .
\end{equation}
On the other hand, the right hand side of (\ref{solva2}) is 
 the integration of the product of (\ref{M}) and (\ref{phi0}).

For simplicity, 
to obtain the explicit form we assume $f_0=1/(1+\tanh(2))$ in (\ref{back}).
In this case the coefficients in (\ref{determine}) are reduced to 
$\rho_{23}=-3/2$, $\rho_{32}=-\beta$, $\rho_{41}=-1/4$, $\eta=1/(4\beta)$
$c_0=2^6 f_0/3^3=1.20689$,  
and $\beta=3\sqrt{3}/(8\sqrt{2}f_0)=
0.902037$, $6c_0/|(UV)'''|=9/4$,
$\theta_+=1.2897187$, $\theta_-=-0.387814$. 
Finally we obtain
\begin{equation}\label{hatten}
\{L-(\theta_+-\theta_-)\}\dot \gamma =4\beta \gamma^2\{
\frac{\theta_+}{\theta_+^2+1}(1-\frac{\gamma}{\gamma_+})
-\frac{\theta_-}{\theta_-^2+1}(1-\frac{\gamma}{\gamma_-})\},
\end{equation}
where we use $\displaystyle\frac{I_{n+1}^{(\pm)}}{I_n^{(\pm)}}=
\displaystyle\frac{2n \theta_{\pm}^2+1}{(2n+1)\theta_{\pm}^2+1}$.
Here $\gamma_{\pm}$ is 
given by	
\begin{eqnarray}\label{determine}
\gamma_{\pm}^{-1}&=&
2+\theta_{\pm}^2\left(2-3\frac{I_2^{(\pm)}}{I_1^{(\pm)}}\right) \nonumber\\
& &+2\eta[ 3\rho_{32}\left(1-\frac{I_2^{(\pm)}}{I_1^{(\pm)}}\right)
+\frac{\rho_{41}}{\theta_{\pm}}
\left(\frac{I_0^{(\pm)}}{I_1^{(\pm)}}-2
+\frac{I_2^{(\pm)}}{I_1^{(\pm)}}\right)-
\rho_{23}\theta_{\pm}\frac{I_2^{(\pm)}}{I_1^{(\pm)}}]. 
\end{eqnarray}
Thus, we obtain $\gamma_{\pm}$ as
\begin{equation}\label{selection}
\gamma_{\pm}=\frac{8\beta\theta_{\pm}(3\theta_{\pm}^2+1)}
{37\theta_{\pm}^4+8\theta_{\pm}^2-8} .
\end{equation}
Substituting (\ref{selection}) into (\ref{hatten}) we obtain
\begin{equation}\label{f-hatten}
(L-\theta_++\theta_-)\dot \gamma
=\alpha\gamma^2(1-\frac{\gamma}{\gamma^*}) ,
\end{equation}
where
\begin{equation}\label{result}
\alpha=\frac{24\beta\sqrt{\beta^2+2}}{4\beta^2+9};
\quad \gamma^*=\frac{3(12\beta^2+25)}{61\beta^2+132}=0.574189\cdots .
\end{equation}
From (\ref{scaling}) the amplitude $A\epsilon$ of $R_0$  can be
regarded as the order parameter of  phase separation, which is given by
\begin{equation}\label{order}
A\epsilon\equiv \frac{3}{2}\epsilon \sqrt{\gamma^{*}}=1.13663 \epsilon.
\end{equation}
In the vicinity of $\gamma^*$ the time evolution of $\gamma$ is
described by
\begin{equation}\label{gamma(t)}
\gamma(\tau)\simeq \gamma^*+A\exp[-\frac{\alpha \gamma^*}{L}\tau]
\end{equation}
Notice that $\gamma^*$ is the stable fixed point in the time evolution 
(\ref{f-hatten}).
where we use $\displaystyle\frac{\theta_+-\theta_-}{L}\ll 1$.

Two  remarks on the result of this section are addressed:
 We recall that to derive 
(\ref{result}) from (\ref{solva2}) we assume $f_0=1/(1+\tanh(2))$.
Although to obtain the result for any $f_0$ is not difficult,
we omit such a generalization to avoid long and tedious calculations.
 Thus, the expression with
$\beta=0$ in (\ref{result}) does not recover the result by Komatsu and 
Sasa\cite{komatsu}. 
As the second remark,
 the characteristic time for the relaxation
(\ref{f-hatten}) or (\ref{gamma(t)}) is proportional to the system size $L$.
This result is reasonable, because the time needed to simulate (\ref{rOV}) 
is proportional to number of cars $N$. This tendency 
has been   confirmed
by our simulation. 

\subsection{Simulation}

To check the validity of our analysis in the previous subsection
we perform the numerical simulation
of (\ref{rOV}) and (\ref{back}) with $f_0=1/(1+\tanh(2))$
near the critical point (\ref{cp})
 under the periodic boundary condition. We adopt 
the classical fourth-order Runge-Kutta scheme
with fixed time interval $\Delta t=2^{-4}$. Since our purpose is 
the quantitative test of (\ref{asymmetry}) and (\ref{result}),
the initial condition is restricted to the localized symmetric
form $r_n= 18.7/N(\tanh(n-N/4)-\tanh(n-3N/4)-1)$ where $N$ is
the number of cars. Taking into account the scaling properties
we perform the simulation for  the set of parameters $(\epsilon, N)=
(1/2, 32), (1/4,64), (1/8,128), (1/16,256)$  until $r_n$ relaxes to
a  steady propagating state. Our  results are plotted in Figs.1 and 2.

Figure 1 displays  points which have the maximum $h_{max}$
and the minimum $h_{min}$ values
of successive car distance in
each parameter set, and theoretical coexistence curve 
\begin{equation}\label{coex}
a=a_c\left(1- \frac{(h-h_c)^2}{A^2}\right); 
\quad A=1.13663\cdots , 
\end{equation} 
where we use $a=a_c(1-\epsilon^2)$ and $A\epsilon=h-h_c$.
The agreement with each other is obvious.
From this figure we can see that 
one of the branches is in the linearly unstable region but
the theoretical curve  recovers the simulation result.   
We stress that the evaluated value of $A$ from simulation at $\epsilon=1/16$
is $1.14273\cdots$. Thus, the deviation between simulation and theory is
only 0.53 $\%$.
\begin{figure}[htbp]
\centerline{
\epsfbox{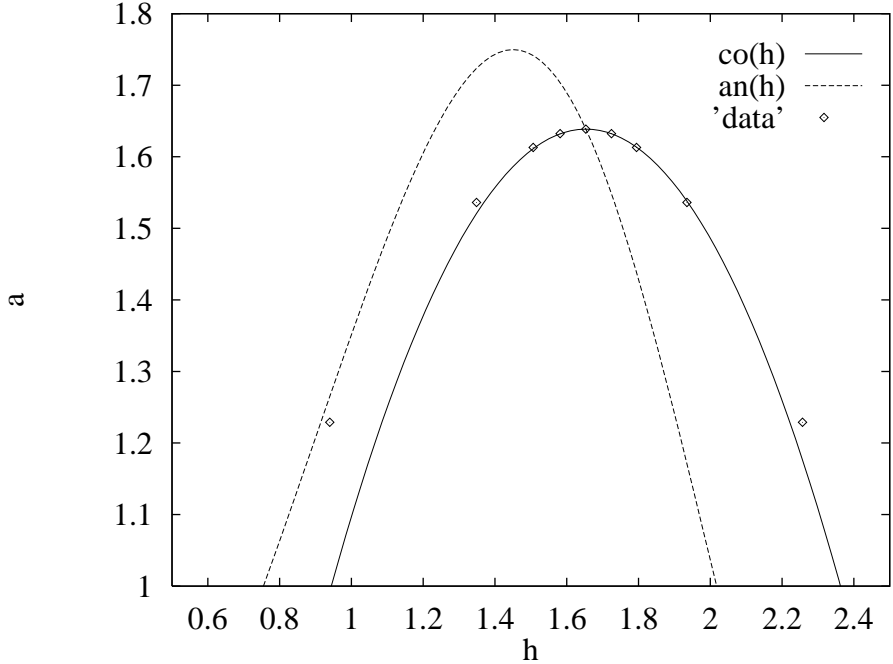}}
\caption{Theoretical coexistence curve (solid line)
$a=a_c(1-(h-h_c)^2/A^2)$ where $A=1.13663$, $h_c=1.65343$ and $a_c=1.63866$,
and the neutral curve (broken line) for the model described by eq.(2)
with $W(h)=\tanh(h-2)+\tanh(2)$, $V(h)=1+(1-\tanh(h-2))/(1+\tanh(2))$
and $f_n=0$.
$h$ denotes the average distance between successive cars.
The data is obtained for
minimum and maximum values of $r_n$ at a given $a$.$^{10)}$}
\label{fig:1}
\end{figure}

\begin{figure}[htbp]
\centerline{
\epsfbox{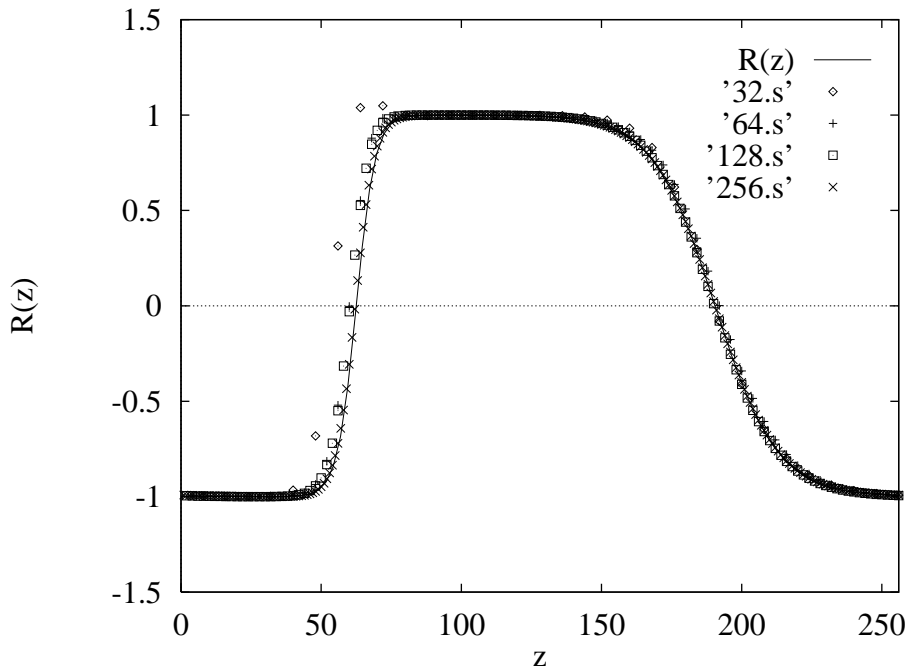}}
\caption{Theoretical curve (solid line) and scaled data obtained from
our simulation for scaled $r_n$. 
Each data denotes
$(\epsilon,N)=(1/2,32),(1/4,64),(1/8,128),(1/16,256)$,
where $\epsilon=(1-a/a_c)^{1/2}$,
$'N.s'$ represents the  data for  $N$ cars (particles).
Theoretical curve is given by
$R(z)=\tanh(\xi \theta_+(z-z_+))-1+\tanh(\xi\theta_-(z-z_-))$
with $\xi=(6 \gamma^*)^{1/2}/16$, $\gamma^*=0.574189$
, $\theta_+=1.2897187$ and $\theta_-=-0.3876814$, where only
$z_+=62.5 $ and $z_-=190.5$ are fitting parameters. 
Spatial scale is measured by
the scale for $N=256$.$^{10)}$}
\label{fig:2}
\end{figure}

Figure 2 demonstrates that the numerical result has a scaling solution
which has an asymmetric kink-antikink pair.
The linear combination of  our theoretical curve
(\ref{asymmetry}) with (\ref{selection}) is plotted as the solid line
by choosing the position of the kink and the anti-kink. Our theoretical 
curve  agrees with the result of our simulation 
 without other fitting parameters.
Thus, we have confirmed the validity of our theoretical analysis.

\section{$f^{-4/3}$ law in power spectra}

The purpose of this section is to clarify the mechanism to appear 
$P(f)\sim f^{-4/3}$ law in power spectra.
We, thus,  extend the one-dimensional 
model (\ref{powder}) 
to a stochastic model (\ref{1d_powder}) supplemented by 
the white noise.
We will demonstrate the simple model reproduces $\alpha = 4/3$  near the 
neutral curve of the linear stability analysis of uniform states.

At first, we briefly summarize the result of linear stability.
Although the framework is common with that in section 3.1,
the explicit forms are a little different.
The neutral curve is given by
$T_n=\tilde U'^2/\varphi''(h)$. 
The critical point is given by the cross point of $\tilde U''(h_c)=0$
on the neutral curve.
We will focus on the behaviors for weakly unstable or stable region
at $T=T_c(1-\mu)$ with $|\mu|\ll 1$.

Let us simulate (\ref{1d_powder})
directly.
Adopting  
\begin{equation}
\tilde U(r)=\tanh(r-2)+\tanh(2),\quad  \varphi(r)={\rm sech}^2(r)
\end{equation}
with  $\zeta=2$,  $N=256$,
$T_c=3.95798\cdots$ ,
and $h=2$ at $t=0$,
we numerically calculate (\ref{1d_powder}) 
by the classical Runge-Kutta method 
 until $t=2^{11}$ with time interval $\Delta t=1/2^4$
under the periodic boundary condition. 
We use 
 the uniform random number distributed between $- X$ and $X$ 
with $X=9/1024$ for $f_n(t)$.
 Figure 3 displays  
the power spectrum $P(f)=<|\tilde \rho(f)|^2>$ 
obtained from our simulation of (\ref{1d_powder}) at $\mu=1/64$, 
where $\tilde \rho(f)$ is the Fourier
transform of  the discretely sampled data of the density 
$\rho(t)=\displaystyle\frac{1}{N}\sum_{n} \frac{1}{r_n(t)}$ 
with the time interval 1.
This clearly supports $P(f)\sim f^{-4/3}$ law 
in the range of $f$ between $10$ and $1000$ 
as in the experiment\cite{chuo}.
It should be noticed 
that the data in Fig.3 may suggest  steeper slope than $f^{-4/3}$ 
in low frequency region $f<10$. Although the steep slope
close to $f^{-3/2}$ at small $f$ 
is not observed
in the experiment by Moriyama et al.\cite{chuo},
it can be explained easily by the diffusive behavior of an interactive
 pair of kink and antikink. 
The details of the process to produce
$f^{-3/2}$  will be reported elsewhere.
  
From the examinations of several  values of $\mu$, 
we have confirmed that the qualitative results
are insensitive to  the sign of $\mu$ when
$|\mu|\ll 1$.  This result is reasonable because near the neutral curve
the time scale of relaxation or growth of fluctuations is much longer than
the  time scale induced by the noise $f_n(t)$.
Our numerical result suggests that the linear relaxation 
theory of fluctuations 
 can be used to explain $P(f)\sim f^{-4/3}$.
This $f^{-4/3}$ law can been seen from the simulation of
traffic flow (\ref{rOV}) with the noise.
\begin{figure}[htbp]
\centerline{
\epsfbox{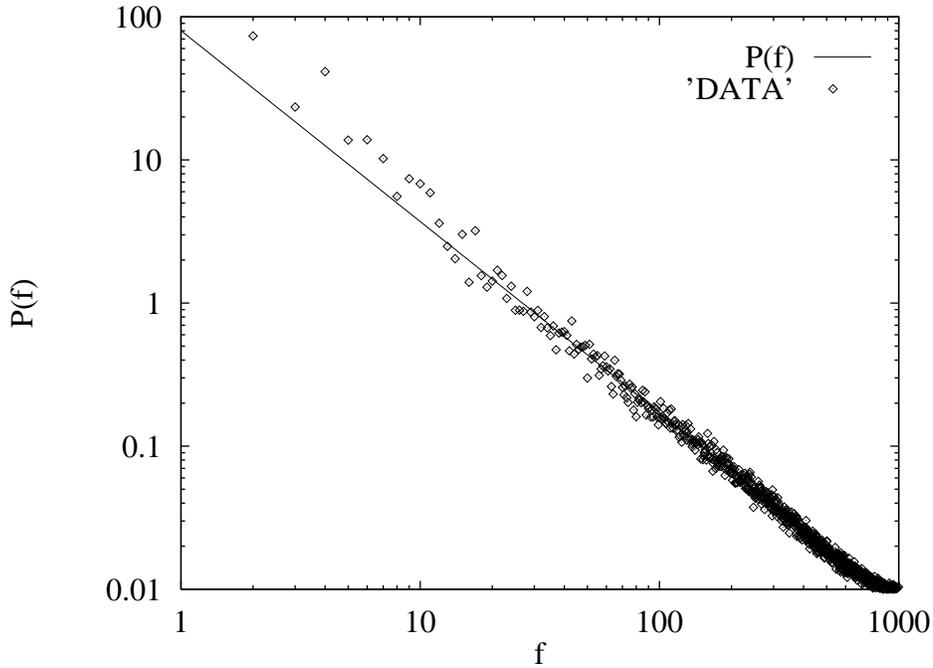}}
\caption{Log-log plot of power spectrum $P(f)$ obtained from
the simulation of eq.(2.7) with (4.1), where the unit of $f$ is $1/(2\pi)$
and the unit of $P(f)$ is not normalized. 
The solid line represents $f^{-4/3}$. }
\label{fig:3}
\end{figure}

Thus, let us briefly explain how to appear $f^{-4/3}$ law from
the  behavior of structure factor   
\begin{equation}\label{scattering}
S_k(t)\equiv
\sum_{n,m}<\exp[i k(  r_n(t)-  r_m(0))]> 
= 
\frac{1}{N}\sum_{n, m}\exp \left[-\frac{k^{2}}{2}\phi_{nm}(t)\right],
\end{equation}
where
$\phi_{nm}(t)=<( r_n(t)- r_m(0))^2>$.
Notice that the structure factor is 
directly related to the auto-correlation function
as $S_k(t)=\int_0^tdt'<\rho_k(t+t')\rho_{-k}(t)>$,
where $\rho_k(t)$ is 
 the Fourier component of the density field of particles $\rho({\bf r},t)$.
In weakly stable states, 
{\it i.e.} $\mu<0$ and $|\mu|\ll 1$,  $S_k(t)$ can be  
calculated
 as in the case of polymer dynamics\cite{com}.
With the aid of the expansion of $\sigma_+$, 
all of models  are
reduced to
\begin{equation}\label{4}
\partial_{\tau}r(z,\tau)-\partial_z^3 r(z,\tau)=\epsilon [\partial_z^2-\partial_z^4]
r(z,\tau)+\xi(z,\tau) 
\end{equation}
in weakly stable region.
When we start from (\ref{1d_powder}) the scaled variables are
given by  $\tau=\epsilon^3\beta t$, 
$z=\displaystyle\frac{2\zeta}{3c_0}\epsilon (x+c_0t)$,
$\xi(z,\tau)=\epsilon^3 \beta f_n(t)$ with 
$\epsilon=\displaystyle\frac{3\sqrt{c_0}}{\zeta}\sqrt{-\mu}$ and
 $\beta=\displaystyle\frac{4}{3\sqrt{c_0}}$.  
The solution of  (\ref{4}) 
is given by
\begin{equation}\label{r_q}
 \tilde r_k(\tau)\simeq \int_{0}^{\tau}ds
\exp[\lambda_k(\tau-s)]\tilde \xi_k(s),
\end{equation}
where $\lambda_{k}=i k^3 -\epsilon k^2(1+k^2)$.
Thus, we obtain the correlation
\begin{equation}\label{r-r}
< \tilde r_k(\tau) \tilde r_{-k}(0)>
=\frac{D}{2\epsilon l k^2(1+k^2)}
\exp[\lambda_k\tau],
\end{equation}
where $l$ is the system size in this unit, and 
we use $<\tilde \xi_k(\tau)\tilde \xi_p(\tau')>=
\displaystyle\frac{D}{l}\delta_{k+p,0}\delta(\tau-\tau')$.

Substituting (\ref{r-r}) into 
$\phi(z,z',t)=<(r(z,t)-r(z',t))^2>$
which is the continuous limit of the scaled $\phi_{nm}$,
we obtain
\begin{equation}
\label{phi3}
\phi(z,z',t)=2D_G t+\frac{D}{2\epsilon l}\sum_{n\ne 0}
\frac{1}{k^2(1+k^2)} 
\left\{
|e^{ik z}-e^{ik z'}|^2+2(1-e^{\lambda_k t})e^{ik(z-z')}\right\}.
\end{equation}
With the aid of 
$\displaystyle\sum_{n\ne 0}\displaystyle\frac{1}{k^2(1+k^2)}
\simeq \displaystyle\frac{l^2}{3}-l$ and
$\displaystyle\sum_{n\ne 0}\displaystyle\frac{\cos(k(z-z'))}{k^2(1+k^2)}
\simeq \frac{l^2}{3}-l|z-z'|$, the first term in the summation in
 eq.(\ref{phi3}) is
reduced to
\begin{equation}\label{sum1}
\sum_{n\ne 0}
\displaystyle\frac{|e^{ik z}-e^{ik z'}|^2}{k^2(1+k^2)}
\simeq 2l|z-z'|.
\end{equation}
On the other hand, the second term in the summation in eq.(\ref{phi3}) becomes
 \begin{eqnarray}\label{sum2}
\sum_{n\ne 0}
\frac{e^{ik(z-z')}}{k^2(1+k^2)}(1-e^{\lambda_k t})
&=&2\sum_{n=1}^{\infty} \frac{\cos[k(z-z')]}{k^2(1+k^2)}
 \{1-\cos(k^3t)\} \nonumber \\
& & +2\sum_{n=1}^{\infty} \frac{\sin[k(z-z')]}{k^2(1+k^2)}\sin(k^3t),
\end{eqnarray}
where we use the approximation $\lambda_k\sim i k^3$.
Replacing the summation $\sum_{n=1}^{\infty}$ to the integral 
$\int_0^{\infty}dk$,
and substituting (\ref{phi3})-(\ref{sum2}) into (\ref{scattering}),
 we obtain
\begin{equation}\label{form}
S_k(\tau)\simeq 
2\int_{0}^l dw \exp[-D_G k^2\tau-\frac{D k^2}{2\epsilon}w -
\frac{D k^2}{\pi \epsilon}\tau^{1/3} h(u)], 
\end{equation}
where $w=|z-z'|$, $u=x\tau^{-1/3}$, and 
the argument of $S_k$ is replaced by the scaled time. 
$D_G$ is the diffusion constant for the center of mass
in (\ref{4}),  and $h(u)$ is
\begin{equation}\label{h(u)}
h(u) = \int_{0}^{\infty}dQ[\frac{\cos (Qu)}{Q^{2}(1+\tau^{-2/3}Q^2)}
(1-\cos (Q^3)) 
+ \frac{\sin(Qu)\sin(Q^3)}{Q^2(1+\tau^{-2/3}Q^2)}], 
\end{equation}
where $Q^{3}=k^{3}\tau$, 
and $w = |z-z'|$.

In the long time limit, eq.(\ref{form}) is governed by
the diffusion of the center of mass. 
For wide range of time, however, 
the contribution from the second and the third terms 
in (\ref{form}) are dominant because of $1/\epsilon\gg 1$.
In such the case, the first term is negligible, and
$h(u)$  converges to 
\begin{equation}\label{h(0)}
h(0)=\int_0^{\infty}dQ\frac{1-\cos(Q^3)}{Q^2}=\int_0^{\infty}dx
\frac{\sin x}{x^{1/3}}=\frac{\pi}{\Gamma(1/3)} ,
\end{equation} 
 as time goes on.   
From
 $\lim_{l\to \infty} \int_0^l dw \exp[-Dk^2 w/(2\epsilon)]=
2\epsilon/Dk^2$, we obtain
\begin{equation}\label{result_sk}
S_k(\tau)\simeq \frac{4\epsilon}{D k^2}
\exp[-\frac{D k^2}{\epsilon\Gamma(1/3)}\tau^{1/3}]
\end{equation}
in intermediate time range.
In the limit of small $\tau$, 
 $S_k(\tau)\propto 1- 
\displaystyle\frac{D k^2}{\epsilon\Gamma(1/3)}k^2\tau^{1/3}+\cdots$. 
Thus its 
Fourier transform, which is nothing but the power spectrum 
$P_k(f)=<|\tilde \rho_k(f)|^2>$
obeys
\begin{equation}\label{spectrum}
P_k(f)\sim f^{-\alpha}, \quad \alpha = 4/3 \quad 
({\rm as}\quad f \to \infty),
\end{equation}
where use was made of 
$\int_{-\infty}^{\infty} d\tau e^{i 2\pi f \tau}|\tau|^{1/3}\propto f^{-4/3}$.
The value $4/3$ is identical to the one obtained by 
the experiment\cite{chuo} and 
numerical simulations\cite{chuo,peng94}.  Thus our model 
(\ref{1d_powder}) reproduces $\alpha = 4/3$.
This result 
should be valid even 
when we start 
from fluid models\cite{KK,batchelor,sasa,komatsu93}
 since the result is determined by the universal 
feature near the neutral curve as in (\ref{4}). 
It should be noted that the appearance 
of this power-law form in the original model (\ref{1d_powder}) is only for 
$f < \zeta$ since we eliminate the fast decaying mode $\sigma_{-}$ in our 
analysis.  This tendency is also observed as the higher-frequency cutoff 
in the experiment\cite{chuo}.
Thus, $f^{-4/3}$ law is determined by short time behavior of 
the  dynamics of density waves induced by the noise,
which  
is essentially determined by the 
linear dispersion relation $\lambda_k\sim i k^3$.

  In this section we have confirmed the universal law 
$P(f)\sim f^{-4/3}$of  in the frequency spectrum of density correlation 
function from  both the simulation and the theory. 
We have also clarified the mechanism to emerge $f^{-4/3}$ spectrum
which is related to the critical slowing down of the density fluctuations.
 It should be noticed that the continuous increase of
$\alpha$ in LGA\cite{peng95} from $\alpha=0$ to 2
with the particle density is consistent with 4/3 law and our picture,
 because the spectrum determined by the noise
in linearly stable uniform state far from the neutral curve
should be white ($\alpha=0$) and
 the effective exponent of the power-law
becomes large when the exponential decay (i.e. $\alpha=2$)
in the off-critical region exists.
There is, however,
discrepancy between our results with the one on the experiment in liquids
\cite{nakahara97}.  The reason of this difference should be clarified in
the future.

\section{Concluding Remarks}

As we have seen in section 3, 
our theoretical analysis gives very precise results
on the phase separation between jam and non-jam phases
for pure one-dimensional models.
Of course, we do not think that our analysis is perfect.
Since, for example,  we omit the time evolution of reduced dynamical models,
 we cannot explain the reason why the linearly unstable
branch of the coexistence curve is stable in simulation (see Fig.1).
The validity of
choices of (\ref{zero-pbc}) and (\ref{psi0}) are also not confirmed
from mathematical point of views, although these choices work very well.
To clarify the above points will be future subject of the research.

Let us comment on the universality class of traffic flows and 
granular flows. 
All of models introduced here except for (\ref{OV}) have
asymmetric kink-antikink pairs and qualitatively resemble 
behaviors with each other.
In fact, we\cite{wada} have already checked the quantitative
validity of
our methods presented here for the fluid model in traffic flows\cite{KK}.
The results are almost identical to those explained in section 3.
On the other hand, OV model in (\ref{OV}) which is a special case of
the above generalized models
loses some universal properties.
It should be noticed that the limitation of OV model
 has been suggested by Komatsu and
Sasa\cite{komatsu} (see the last part in their paper).
Therefore, we believe that our analysis is meaningful to characterize
universal feature of
 one-dimensional dissipative flows such as granular flows  and
traffic flows.

On the other hand, we have extended a one-dimensional deterministic model
to a stochastic model supplemented by the white noise.
This model clearly reproduces $P(f)\sim f^{-4/3}$ law
as in the experiment\cite{chuo}.
We also give a simple argument for the reason why we obtain $f^{-4/3}$ law.
Readers may be skeptical whether our one-dimensional models
supplemented by the white noise can describe true behavior of
quasi one-dimensional systems in spite of good agreement with
the experiment.
To reply such the question, we have already introduced a simple
model for two-lanes traffic flows\cite{takaai}, where 
the vehicles have spin variable to specify what lane is chosen
and lane-change of vehicles
 is governed by Glauber dynamics for anti-ferro-magnetism.
Our preliminary result of simulation also supports $f^{-4/3}$ law.
Thus, $f^{-4/3}$ law is believed to be 
universal without regard to the choice of models.
The effect of nonlinearity in the theoretical argument
and the validity of the introduction  of
the Gaussian white noise in (\ref{1d_powder}) 
will be discussed elsewhere.

In conclusion, we have 
proposed a simple generalized optimal velocity model
(\ref{rOV}). 
Based on the perturbation analysis of asymmetric kink solution (\ref{asymmetry})
we obtain the selected values of its amplitude, propagating velocity
and width of kink as in (\ref{result}).
 The accuracy 
and relevancy of the solution has been confirmed by the direct simulation.
We also extend a one-dimensional model to a quasi one-dimensional model
with adding the white noise. From the simulation of the noise sustained model
and analytic linear relaxation theory we have confirmed the universality
of $f^{-4/3}$ law in power spectra in density auto-correlation function.

\section*{Acknowledgments}

One of the authors (HH) thanks S. Wada, T.Takaai and 
O. Moriyama for fruitful discussion.
This work is partially supported by Grant-in-Aid of Ministry of Education,
Science and Culture of Japan (09740314).




\begin{thebibliography}{99}
\bibitem{comment}
In other words, the contribution from the term $\rho_{41}$ exists if
the system is under the free boundary condition\cite{ken2}.
\end{thebibliography}
\end{document}